\documentstyle[twoside,epsf,psfig]{article}

\def\be{\begin{equation}}
\def\ee{\end{equation}}

\catcode`\@=11
\long\def\@makefntext#1{
\protect\noindent \hbox to 3.2pt {\hskip-.9pt  
$^{{\eightrm\@thefnmark}}$\hfil}#1\hfill}		%CAN BE USED 

\def\@makefnmark{\hbox to 0pt{$^{\@thefnmark}$\hss}}	%ORIGINAL 
	
\def\ps@myheadings{\let\@mkboth\@gobbletwo
\def\@oddhead{\hbox{}
\rightmark\hfil\eightrm\thepage}   
\def\@oddfoot{}\def\@evenhead{\eightrm\thepage\hfil
\leftmark\hbox{}}\def\@evenfoot{}
\def\sectionmark##1{}\def\subsectionmark##1{}}

\oddsidemargin=\evensidemargin
\newcounter{sectionc}\newcounter{subsectionc}\newcounter{subsubsectionc}
\renewcommand{\section}[1] {\vspace{12pt}\addtocounter{sectionc}{1} 
\setcounter{subsectionc}{0}\setcounter{subsubsectionc}{0}\noindent 
	{\tenbf\thesectionc. #1}\par\vspace{5pt}}
\renewcommand{\subsection}[1] {\vspace{12pt}\addtocounter{subsectionc}{1} 
\setcounter{subsubsectionc}{0}\noindent 
{\bf\thesectionc.\thesubsectionc. {\kern1pt \bfit #1}}\par\vspace{5pt}}
\renewcommand{\subsubsection}[1] {\vspace{12pt}\addtocounter{subsubsectionc}{1}
	\noindent{\tenrm\thesectionc.\thesubsectionc.\thesubsubsectionc.
	{\kern1pt \tenit #1}}\par\vspace{5pt}}

\newcounter{appendixc}
\newcounter{subappendixc}[appendixc]
\newcounter{subsubappendixc}[subappendixc]
\renewcommand{\thesubappendixc}{\Alph{appendixc}.\arabic{subappendixc}}
\renewcommand{\thesubsubappendixc}
	{\Alph{appendixc}.\arabic{subappendixc}.\arabic{subsubappendixc}}

\renewcommand{\appendix}[1] {\vspace{12pt}
        \refstepcounter{appendixc}
        \setcounter{figure}{0}
        \setcounter{table}{0}
        \setcounter{lemma}{0}
        \setcounter{theorem}{0}
        \setcounter{corollary}{0}
        \setcounter{definition}{0}
        \setcounter{equation}{0}
        \renewcommand{\thefigure}{\Alph{appendixc}.\arabic{figure}}
        \renewcommand{\thetable}{\Alph{appendixc}.\arabic{table}}
        \renewcommand{\theappendixc}{\Alph{appendixc}}
        \renewcommand{\thelemma}{\Alph{appendixc}.\arabic{lemma}}
        \renewcommand{\thetheorem}{\Alph{appendixc}.\arabic{theorem}}
        \renewcommand{\thedefinition}{\Alph{appendixc}.\arabic{definition}}
        \renewcommand{\thecorollary}{\Alph{appendixc}.\arabic{corollary}}
        \renewcommand{\theequation}{\Alph{appendixc}.\arabic{equation}}
        \noindent{\tenbf Appendix \theappendixc #1}\par\vspace{5pt}}
\newcommand{\subappendix}[1] {\vspace{12pt}
        \refstepcounter{subappendixc}
        \noindent{\bf Appendix \thesubappendixc. {\kern1pt \bfit #1}}
	\par\vspace{5pt}}
\newcommand{\subsubappendix}[1] {\vspace{12pt}
        \refstepcounter{subsubappendixc}
        \noindent{\rm Appendix \thesubsubappendixc. {\kern1pt \tenit #1}}
	\par\vspace{5pt}}

\newcommand{\textlineskip}{\baselineskip=13pt}
\newcommand{\smalllineskip}{\baselineskip=10pt}

\def\abstracts#1#2#3{{
	\centering{\begin{minipage}{4.5in}\footnotesize\baselineskip=10pt
	\parindent=0pt #1\par 
	\parindent=15pt #2\par
	\parindent=15pt #3
	\end{minipage}}\par}} 

\def\keywords#1{{
	\centering{\begin{minipage}{4.5in}\footnotesize\baselineskip=10pt
	{\footnotesize\it Keywords}\/: #1
	 \end{minipage}}\par}}

\renewenvironment{thebibliography}[1]
        {\frenchspacing
	 \ninerm\baselineskip=11pt
         \begin{list}{\arabic{enumi}.}
        {\usecounter{enumi}\setlength{\parsep}{0pt}     
	 \setlength{\leftmargin 12.7pt}{\rightmargin 0pt}%FOR 1--9 ITEMS
         \setlength{\itemsep}{0pt} \settowidth
	{\labelwidth}{#1.}\sloppy}}{\end{list}}

\newcounter{itemlistc}
\newcounter{romanlistc}
\newcounter{alphlistc}
\newcounter{arabiclistc}

\newcommand{\fcaption}[1]{
        \refstepcounter{figure}
        \setbox\@tempboxa = \hbox{\footnotesize Fig.~\thefigure. #1}
        \ifdim \wd\@tempboxa > 5in
           {\begin{center}
        \parbox{5in}{\footnotesize\smalllineskip Fig.~\thefigure. #1}
            \end{center}}
        \else
             {\begin{center}
             {\footnotesize Fig.~\thefigure. #1}
              \end{center}}
        \fi}

\newcommand{\tcaption}[1]{
        \refstepcounter{table}
        \setbox\@tempboxa = \hbox{\footnotesize Table~\thetable. #1}
        \ifdim \wd\@tempboxa > 5in
           {\begin{center}
        \parbox{5in}{\footnotesize\smalllineskip Table~\thetable. #1}
            \end{center}}
        \else
             {\begin{center}
             {\footnotesize Table~\thetable. #1}
              \end{center}}
        \fi}

\def\@citex[#1]#2{\if@filesw\immediate\write\@auxout
	{\string\citation{#2}}\fi
\def\@citea{}\@cite{\@for\@citeb:=#2\do
	{\@citea\def\@citea{,}\@ifundefined
	{b@\@citeb}{{\bf ?}\@warning
	{Citation `\@citeb' on page \thepage \space undefined}}
	{\csname b@\@citeb\endcsname}}}{#1}}

\newif\if@cghi
\def\cite{\@cghitrue\@ifnextchar [{\@tempswatrue
	\@citex}{\@tempswafalse\@citex[]}}
\def\citelow{\@cghifalse\@ifnextchar [{\@tempswatrue
	\@citex}{\@tempswafalse\@citex[]}}
\def\@cite#1#2{{$\null^{#1}$\if@tempswa\typeout
	{IJCGA warning: optional citation argument 
	ignored: `#2'} \fi}}

\def\pmb#1{\setbox0=\hbox{#1}
	\kern-.025em\copy0\kern-\wd0
	\kern.05em\copy0\kern-\wd0
	\kern-.025em\raise.0433em\box0}

\def\fnt#1#2{\footnotetext{\kern-.3em
	{$^{\mbox{\scriptsize #1}}$}{#2}}}

\def\fpage#1{\begingroup
\voffset=.3in
\thispagestyle{empty}\begin{table}[b]\centerline{\footnotesize #1}
	\end{table}\endgroup}

\font\tenrm=cmr10
\font\tenit=cmti10 
\font\tenbf=cmbx10
\font\bfit=cmbxti10 at 10pt
\font\ninerm=cmr9

\font\eightrm=cmr8

\def\qed{\hbox{${\vcenter{\vbox{	          %HOLLOW SQUARE
   \hrule height 0.4pt\hbox{\vrule width 0.4pt height 6pt
   \kern5pt\vrule width 0.4pt}\hrule height 0.4pt}}}$}}

\begin{document}
\setlength{\textheight}{7.7truein}    %FOR 2ND PAGE ONWARDS

%\runninghead{Instructions for Typesetting Camera-Ready
%Manuscripts $\ldots$} {Instructions for Typesetting Camera-Ready
%Manuscripts $\ldots$}

\normalsize\textlineskip
\thispagestyle{empty}
\setcounter{page}{1}

%%\copyrightheading{}		%{Vol.~0, No.~0 (1999) 000--000}

\vspace*{0.88truein}

\fpage{1}
\centerline{\bf BACKGROUND ESTIMATION IN A GRAVITATIONAL
WAVE EXPERIMENT}
%\vspace*{0.035truein}
%\centerline{\bf MANUSCRIPTS USING COMPUTER SOFTWARE\footnote{For
%the title, try not to use more than 3 lines. Typeset the title
%in 10 pt Times Roman, uppercase and boldface.}}
\vspace*{0.37truein}
\centerline{\footnotesize PIA ASTONE}
\vspace*{0.015truein}
\centerline{\footnotesize\it INFN, sezione ``La Sapienza'',
Physics Department, P. Aldo Moro, 2}
\baselineskip=10pt
\centerline{\footnotesize\it Rome, I-00185,Italy}
\vspace*{10pt}
\centerline{\footnotesize SERGIO FRASCA}
\vspace*{0.015truein}
\centerline{\footnotesize\it INFN, sezione ``La Sapienza'' and
Physics Department, P. Aldo Moro, 2}
\baselineskip=10pt
\centerline{\footnotesize\it Rome, I-00185,Italy}
\vspace*{10pt}
\centerline{\footnotesize GUIDO PIZZELLA}
\vspace*{0.015truein}
\centerline{\footnotesize\it INFN, Laboratori Nazionali di Frascati and
University of Rome ``Tor Vergata'' Physics Department, 
Via della Ricerca Scientifica}
\baselineskip=10pt
\centerline{\footnotesize\it Rome, I-00133,Italy}
\vspace*{10pt}

\vspace*{0.21truein}
\abstracts{The problem to estimate the background due to accidental
coincidences in the search for coincidences in gravitational
wave experiments is discussed. The use of delayed coincidences
obtained by orderly shifting the event times of one of the
two detectors is shown to be the most correct}{}{}

\vspace*{10pt}

%\textlineskip			%) USE THIS MEASUREMENT WHEN THERE IS
\vspace*{12pt}			%) NO SECTION HEADING

\vspace*{1pt}\textlineskip	%) USE THIS MEASUREMENT WHEN THERE IS

%\Instfoot{xx}{Sezione INFN di Roma 1, Rome, Italy\\
%   {\rm Email}: {\tt pia.astone@roma1.infn.it} \\ 
%   {\rm URL}: {\tt http://grwav3.roma1.infn.it}}
%\Instfoot{yy}{ Universit{\`a} ``La Sapienza'' and
%        Sezione INFN di Roma 1, Rome , Italy\\
%        {\rm Email}: {\tt sergio.frasca@roma1.infn.it}\\ 
%        {\rm URL}: {\tt http://grwavsf.roma1.infn.it/}}
%\Instfoot{zz}{ Universit{\`a} ``Tor Vergata'' and
%        LNF Frascati, Rome , Italy\\
%        {\rm Email}: {\tt guido.pizzella@lnf.infn.it}} 
%
%\begin{titlepage}
%\begin{flushright}
%        \small
%          hep-ph/0002004\\
%         Roma1-10XX\\
%         March 1999
%\end{flushright}
%
%{\large\bf Abstract}
%
%\vspace{.5cm} 
%\end{center}

%\end{titlepage}
%%%%%%%%%%%%%%%%%%%%%%%%%%%%%%%%%%%%%%%%%%%%%%%%%%%%%%%%%%%%%%%%%%%%%%%%%%%%%
%%%%%%%%%%%%%%%%%%%%%%%%%%%%%%%%%%%%%%%%%%%%%%%%%%%%%%%%%%%%%%%%%%%%%%%%%%%%%
% from the dvips manual: put a background `DRAFT' on the page
%\special{!userdict begin 
%/bop-hook{gsave 200 30 translate 65 rotate
%           /Times-Roman findfont 216 scalefont setfont
%           0 0 moveto 0.95 setgray (DRAFT) show grestore}def end}
%
%%%%%%%%%%%%%%%%%%%%%%%%%%%%%%%%%%%%%%%%%%%%%%%%%%%%%%%%%%%%%%%%%%%%%%%%%%%%%
%%%%%%%%pia\setcounter{page}{2}
%\input{parte0}
%\input{parte1}
%\input{bibl}

\section{Introduction}

When searching for coincidences due to short bursts of gravitational
radiation (GW) we are faced with the problem that the coincidences
found at zero delay could be casual.
In order to measure the background due to the accidental
coincidences, the most common procedure adopted since the beginning of
the gravitational wave experiments\cite{weber} consists in shifting
the time of occurrence of the events of one of
the two detectors a certain number of times. The distribution of the
delayed coincidences gives the statistical properties of the
background and allows to estimate, in the case of a coincidence
excess, the probability that the excess was accidental.

In this contribution we shall try to catch the problems which
may arise in this procedure and suggest how to cope with them.

\section{Time delay histogram}

For the sake of simplicity we consider here only coincidences between
pairs of detectors, but all the considerations apply to the
general case on $N$ detectors.
The outputs of the background estimation procedure, obtained by off-timing
techniques, are the ``time delay histograms''.

\begin{figure}[hbt]
 \vspace{9.0cm}
\includegraphics{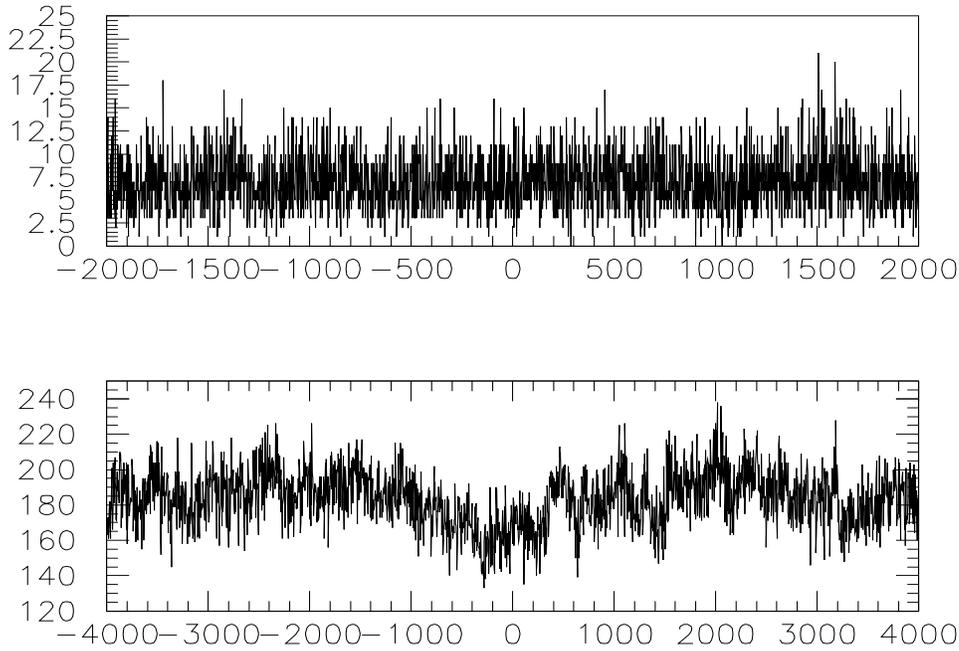}
 \caption{
In the upper figure we show the number of coincidences versus
the time delay in seconds (delay histogram), for EXPLORER/NAUTILUS 1998.
In the lower figure we show the delay histogram for EXPLORER/NIOBE
1995.
        \label{exnani} }
\end{figure}

There are several types of delay histograms. Two typical cases
are shown in Fig.\ref{exnani}. The upper part of the figure
shows the delay histogram obtained with the 1998 IGEC data
of the detectors NAUTILUS\cite{naut} and EXPLORER\cite{expl}.
The lower part of the figure
shows the time delay histogram obtained with
100 days of data recorded by EXPLORER and NIOBE in 1995\cite{astro}.

From the figure we note that the upper histogram can
be considered ``good'', as no particular structures in the data appear.
This lead to the prediction that the distribution of the
accidental coincidences is well described by a
Poisson distribution, as verified in many cases. This reflects the
fact that -in this particular case- over the whole observation time $T_{obs}$
the noise process is stationary or -more specifically- 
the event occurrence times fulfill the conditions which define a Poisson
process.

The lower
figure shows a particular structure, that reflects same non stationary
noise in one or both the detectors. In particular, around the zero delay
the number of off-timing coincidences is clearly systematically lower
compared to the behaviour at $\pm 4000$ s. In this case, the standard
procedure of comparing the $n_c$ coincidences at zero delay with the average
number $\bar{n}$ of shifted
coincidences will lead to underestimate -if any- a physical
effect. On the contrary, suppose,
 even if here it is not the case, $n_c$=210 events.
Given the local background (the background estimated in an interval
$\pm 500$ s, for example) the suspicion may arise that same physical effect
has been observed. The use of the more robust but here meaningless estimation
over $\pm 4000$ s would mask the effect.
If forced, by evident non stationary noise,
to use ``local shifts'' the final estimation will clearly result to be 
less accurate.

\begin{figure}[hbt]
 \vspace{9.0cm}
\includegraphics{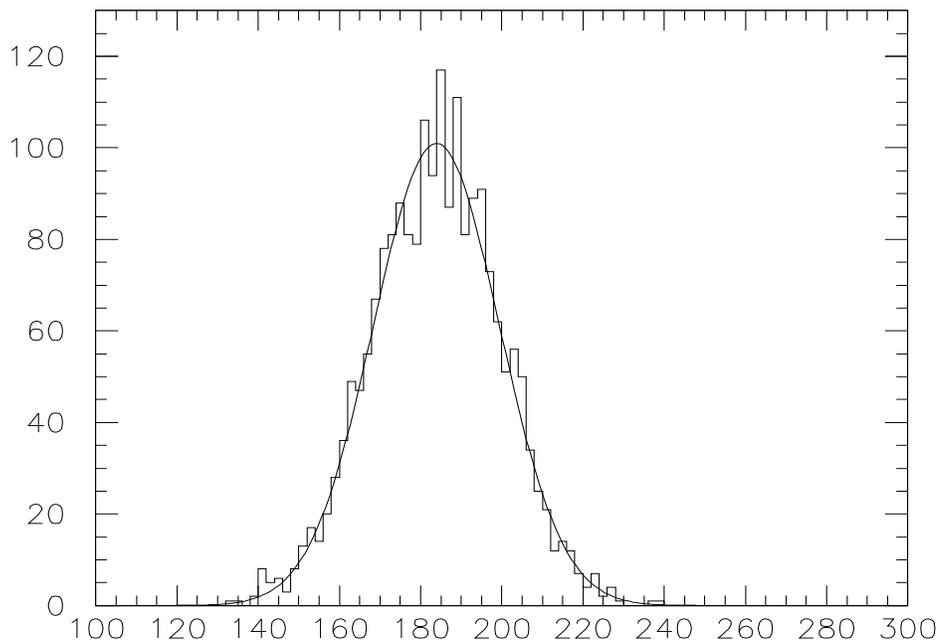}
 \caption{
Distribution of the EXPLORER/NIOBE delayed coincidences.
The line is a gaussian fit.
        \label{CORFDISTRI} }
\end{figure}

One could expect that the non-stationary noise would
give a non-poissonian distribution of the delayed coincidences.
Instead we find, for the data of the lower
part of Fig.\ref{exnani}, the distribution shown in Fig. \ref{CORFDISTRI},
well fitted with a Gaussian curve.
 
This is because the number of the background coincidences
at the different delays
here is so high that (due to the central limit theorem) the observed
distribution is a Gaussian one.

Therefore we cannot use, in general, the distribution of the delayed
coincidences to validate a statistical result.
The natural suggestion is that a coincidence result should always
be presented as a time delay histogram plus the number $n_c$ of
coincidences and the average $\bar{n}$ of the delayed coincidences. 

\section{Remarks on the previous examples and the use of the moving
threshold to select the events}

Going into a more detailed study of the data we noticed that the effect
in Fig. \ref{exnani} is not due to ``stop runs'' in the data
(see the discussion on this point in section 5). %%%%%%%\ref{holes}
It is due to the non stationary noise, which produces in a detector
a highly varying number of event per hour. In the upper part of
Fig.\ref{exnani} for both detectors, EXPLORER and NAUTILUS, 
we had made use of a moving energy threshold adapted to the noise
that keeps nearly constant
the event rate. In the lower part of Fig.\ref{exnani} we had made
use of NIOBE events obtained with a fixed threshold, that produced
an event rate from a few events up to two hundred events per hour.
Thus the use of a moving threshold adapted to the noise is recommended.

The use of a moving threshold reduces the effect of the non 
stationary noise, but we still have a problem when the detectors have
very different sensitivities. This has to be considered, if possible,
when comparing the two events lists.
We use here the terms ``events'' to indicate the quantities measured by the
detectors and ``signals'' to indicate the physical quantities we aim to infer
(see ref.\cite{pizzella}).  
The Explorer and
Nautilus detectors during 1998 had very different noise.
Indicating the noise with the effective temperature $T_{eff}$
\footnote[1]{$T_{eff}$ is a parameter
that is related to the event amplitude $h$ by a simple equation
(i.e. 10 mK means $h=8 \cdot 10^{-19}$).}, 
only a very small number of hours of Explorer have a sensitivity better than
15 mK, which is the worst Nautilus sensitivity. 
This means that the sensitivity of the
global analysis is set by Explorer.

In general, we do not make assumptions on the signals amplitudes and the
standard analysis is  done using all the data. 
Our remark is intended to note that it is worth, in addition
to the standard analysis, to do separate analyses,
considering the different detectors sensitivities and the possible signals
amplitudes. 
If the signal amplitudes are expected to be so large,
compared to the noise of the worst detector, that its detection
efficiency (the fraction of events detected at a given level) 
be $\epsilon \simeq 1$, then 
the standard analysis
can be applied without particular care. 

But, as it usually seems the case, when
the signal amplitudes are expected to be ``small'', then proper additional
analyses must be done using only the data corresponding to similar
sensitivities. Clearly this will reduce the observation time $T_{obs}$, but
the combination of data which are measuring
different physical effects may produce an artificially spoiled 
result\cite{inferring}.

\section{Random coincidences to estimate the background}
The procedure so far described uses ``shifts'' to estimate the background.
One might envisage alternative procedures, such as a random reshuffling
of the times of one of the two sequences. Then the events will
be distributed in a random way over the entire $T_{obs}$.

The same arguments we used in the previous section and, in particular,
the lesson we learn with Fig.\ref{exnani}, show eloquently
what happens in such a case.
In fact, while with the shift procedure we maintain the
data structure and so we can do considerations and derive conclusions
from it, when we have randomly reshuffled the times important information
contained in them is lost forever. We can loose a genuine effect or we
can claim for a possible -false!- discovery.

\begin{figure}[hbt]
 \vspace{9.0cm}
\includegraphics{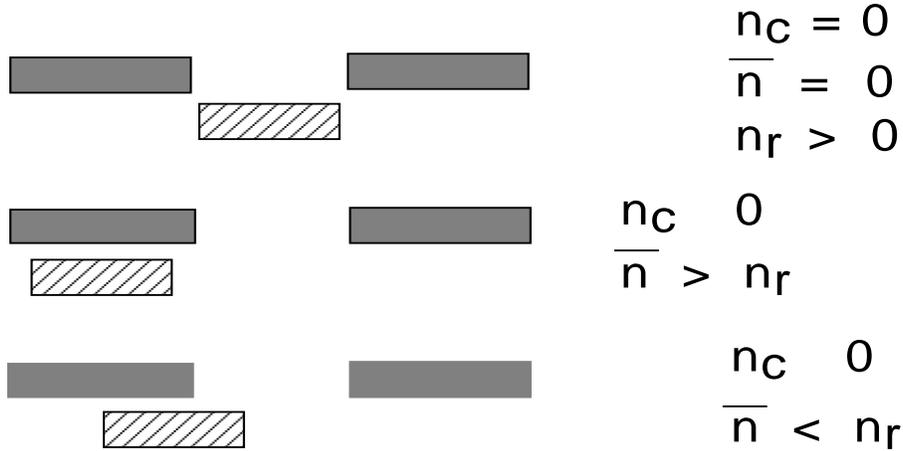}
 \caption{
The time runs horizontally. The black regions indicate the 
event coverage of the first detector, the diagonal marked regions
indicate the event coverage of the second detector.
The upper case occurs when the events of the second detector fall in a
hole of the first detector. In this case we have $n_c=0$, $\bar{n}=0$
for the average estimated with the shifts and $n_r>0$ for the
average estimated with randomly changing the event times.
The other two cases clearly follow.
        \label{hole} }
\end{figure}

It is easy to convince ourselves that the result of a random reshuffling may
overestimate or underestimate the true background,
depending on the relative positions of holes in the two data streams.
This is easily illustrated in Fig.\ref{hole}

\section{The presence of ``stop runs'' in the data}
\label{holes}

Usually the data of both the sequences will contain ``holes'', that
is missing data due to ``stop runs''.
In total the events of the two detectors will cover a common
period of time $T_{obs}$. When we apply the time shift procedure for the
determination of the accidental coincidences the observation time
will be different for each shift, since the events of the detector
with the shifted times might  overlap with a hole in the event list
of the other detector (or viceversa).

If the holes for each detector are randomly
distributed, this change in the observation period
turns out to be negligible, since the decrease of it due to a hole
will be compensated by the increase due to another hole.
In any case it is always possible to estimate this change and normalize
the numbers of accidental coincidences to the real observation
time $T_{obs}$.
This normalization is important if the holes are not randomly
distributed, but always occur for the two detectors at the same times.

\section{Conclusion}
To summarize, the final recommendation is:
\begin{itemize}
\item
to use the ``shifts'', to maintain the information on the noise structure

\item
not to use ``random'' data reshuffling to estimate the background

\item
always give as final result the time delay histogram, plus $n_c$
and $\bar{n}$

\item
use with care detectors with different sensitivities.  

\item
start/stop times are necessary to take into account different periods of
overlapping when shifting

\item
use the adapted threshold to select the events, to make the events occurrence
more stationary

\end{itemize}
%%%%%%%


\begin{thebibliography}{99}
\bibitem{weber}J. Weber, Phys. Rev. Lett. 22, 1320 (1969)
\bibitem{naut}P. Astone et al, Astroparticle Physics, 7 (1997) 231-243 
\bibitem{expl}  P. Astone et al., Phys. Rev. D. 47, 362 (1993).
\bibitem{astro}P.Astone et al, Astroparticle Physics 10 (1999)83-92
\bibitem{pizzella}P.Astone, S.D'Antonio and G.Pizzella, 
This Proceedings (1999)
\bibitem{inferring} P.Astone and G.D'Agostini, CERN-EP/99-126 and 
{\tt hep-ex/9909047}

%\section{The Main Text}
\noindent


\end{thebibliography}
\end{document}